\documentclass[aps,twocolumn,prx,preprintnumbers,showpacs,amsmath,amssymb,superscriptaddress]{revtex4-1}
\usepackage[dvipdfm, colorlinks, citecolor=blue]{hyperref}
\usepackage{graphicx}
\usepackage{dcolumn}
\usepackage{threeparttable}
\usepackage{multirow}
\usepackage{booktabs}
\usepackage{txfonts}
\usepackage{xcolor}
\usepackage{bm}
\usepackage{amssymb}
\usepackage{amsmath}
\usepackage{latexsym}
\usepackage{epsfig}
\usepackage{amsbsy}
\usepackage{array}
\usepackage{tabularx}

\newcommand{\bra}[1]{\langle #1 |}
\newcommand{\ket}[1]{| #1 \rangle}
\newcommand{\ii}{\mathrm{i}}
\newcommand{\ee}{\mathrm{e}}

\newcommand{\qq}{\mathbf{q}}
\newcommand{\kk}{\mathbf{k}}
\newcommand{\nk}{{n\mathbf{\kk}}}

\newcommand{\rr}{\mathbf{r}}
\newcommand{\sg}{\mathbf{g}}

\newcommand{\lG}{\mathbf{G}}
\newcommand{\RR}{\mathbf{R}}
\newcommand{\UG}{\overline{G}}
\newcommand{\OG}{\underline{G}}
\newcommand{\US}{\overline{\Sigma}}
\newcommand{\OS}{\underline{\Sigma}}
\newcommand{\sandw}[2]{\langle #1 | #2 \rangle }
\newcommand{\tpsi}{\tilde\psi}
\newcommand{\pp}{\tilde{p}}
\newcommand{\ppsi}{\tilde\psi}

\begin{document}

\title [] {Cubic scaling $GW$: towards fast quasiparticle calculations}

\author{Peitao Liu}
\email{peitao.liu@univie.ac.at}
\affiliation{University of Vienna, Faculty of Physics and Center for
Computational Materials Science, Sensengasse 8/12, A-1090 Vienna, Austria}
\affiliation{Institute of Metal Research, Chinese Academy of Sciences, Shenyang 110016, China}

\author{Merzuk Kaltak}
\affiliation{University of Vienna, Faculty of Physics and Center for
Computational Materials Science, Sensengasse 8/12, A-1090 Vienna, Austria}

\author{Ji\v{r}\'{\i} Klime\v{s}}
\affiliation{J. Heyrovsk\'{y} Institute of Physical Chemistry, Academy of Sciences of the Czech Republic,
Dolej\v{s}kova 3, CZ-18223 Prague 8, Czech Republic}
\affiliation{Department of Chemical Physics and Optics, Faculty of Mathematics and Physics,
Charles University, Ke Karlovu 3, CZ-12116 Prague 2, Czech Republic}

\author{Georg Kresse}
\affiliation{University of Vienna, Faculty of Physics and Center for
Computational Materials Science, Sensengasse 8/12, A-1090 Vienna, Austria}


\begin{abstract}
Within the framework of the full potential projector-augmented wave methodology,
we present a promising low-scaling $GW$ implementation. It allows for quasiparticle calculations
with a scaling that is cubic in the system size and  linear  in  the number of $k$ points used to
 sample the Brillouin zone. This is achieved by calculating the
 polarizability and self-energy  in real space and imaginary time.
 The transformation from the imaginary time to the frequency domain is
 done by an efficient discrete Fourier transformation with only a
 few nonuniform grid points. Fast Fourier transformations are used to go from
 real space to reciprocal space and vice versa. The analytic continuation from the imaginary to
 the real frequency axis is performed by exploiting Thiele's reciprocal difference approach.
 Finally, the method is applied successfully to predict the quasiparticle energies and
 spectral functions of typical semiconductors (Si, GaAs, SiC, and ZnO),
 insulators (C, BN, MgO, and LiF), and metals (Cu and SrVO$_3$). The results
 are compared with conventional $GW$ calculations. Good agreement is achieved,
 highlighting the strength of the present method.
\end{abstract}

\maketitle

\section{Introduction }\label{Section: intro}
Kohn-Sham (KS) density functional theory (DFT)  in the local density approximation (LDA)~\cite{DFT1965}
 has been proven to be successful in describing the ground-state 
 properties for many weakly correlated materials. However, it fails to predict excited-state
properties~\cite{Aulbur2000}. For instance, DFT within the LDA or generalized gradient approximation (GGA)
always gives a smaller band gap than the experimental value.
This is not surprising, because DFT is just a ground-state theory and there is no formal
justification to interpret DFT eigenvalues as quasiparticle (QP) energies.
In contrast, the $GW$ approximation of  Hedin~\cite{Hedin1965,Hedin1969}
has been widely and successfully applied to the
calculations of QP energies for many kinds of systems (for reviews, see Refs.~\cite{Aryasetiawan2000,Onida2002}),
because it provides a good approximation for the electron's self-energy by
including many-body effects in the electron-electron interaction.
This is achieved by screening the bare exchange interaction
with the inverse frequency-dependent dielectric function.
Moreover, since the $GW$ self-energy can be diagrammatically formulated
in the same many-body framework as dynamical mean field theory (DMFT)~\cite{Georges1996,Kotliar1996},
the $GW$ approximation  not only enables an elegant combination with DMFT,
i.e.,  $GW$+DMFT~\cite{Biermann2003, Held2007}, but also overcomes
the fundamental double counting problems occurring in LDA+DMFT,
because for $GW$+DMFT one actually knows which Feynman diagrams are counted twice~\cite{GW+DMFT2011}.

However, conventional $GW$ calculations are usually restricted to small systems and few $k$ points.
This is related to the fairly high computational cost, which is caused by the evaluation of
the computationally demanding polarizability and
self-energy at a set of real frequencies.
Direct evaluation of the polarizability using the Adler and Wiser formula~\cite{Adler1962, Adler1963}
involves a summation over all pairs of occupied and unoccupied states, leading to the
quartic scaling in the system size and quadratic scaling
in the number of $k$ points used to sample the Brillouin zone (BZ).
Furthermore, calculation of the self-energy for all occupied states in reciprocal space and real frequency is at
least two times more expensive than the evaluation of the polarizability.
To reduce the computational effort, the space-time method~\cite{Godby1995,Godby2000},
which calculates the polarizability and self-energy in real space and  imaginary time,
was proposed. Nevertheless, the space-time method demands considerable storage for the
Green's function and self-energy due to the huge number of real-space grid points.
In addition, to obtain reasonable convergence, fairly dense imaginary-time grids are required.

To circumvent the large storage requirement of the space-time method, a promising scheme
has been recently proposed by Kaltak \emph{et al.}~\cite{KaltakJCTC2014, KaltakPRB2014}.
It allows to calculate the random phase approximation (RPA) correlation energy with a cubic scaling in the
system size and a linear scaling in the number of $k$ points used to sample the BZ.
As in the work of Rojas \emph{et al.}~\cite{Godby1995}, this is achieved by calculating the polarizability
 in real space and imaginary time via contraction over the Green's functions of occupied and
 unoccupied states. The transformation of the polarizability from the imaginary time
 to the frequency domain is performed by an efficient discrete Fourier transformation with only a
 few nonuniform grid points~\cite{KaltakJCTC2014}. Spatial fast Fourier transformations (FFT) within a supercell
are utilized to go from  real space to reciprocal space and vice versa~\cite{KaltakPRB2014}.

Here, we extend Kaltak's scheme~\cite{KaltakJCTC2014, KaltakPRB2014} to
QP calculations in the $GW$ approximation, in which the screened Coulomb interaction
$W$ is calculated within the RPA and the self-energy is
efficiently evaluated via contraction over the Green's function
and $W$  in real space and imaginary time.
Similar spatial FFT as discussed in Ref.~\cite{KaltakPRB2014} are employed, whenever
transformations between the real and reciprocal space are required.
 To transform the self-energy from the imaginary time to the frequency domain,
 nonuniform cosine and sine transformations are used for the even and
 odd parts of the self-energy, respectively. Given that DMFT is usually formulated
on the imaginary frequency axis as well~\cite{Georges1996,Kotliar1996},
our method provides a natural interface for the combination of $GW$ with DMFT.

In this paper, we focus only on the $GW$  QP calculations.
Detailed formulations for our low-scaling $GW$ implementation within the
framework of the projector-augmented wave (PAW) are given.
Considering that the Green's functions
and self-energies in the present implementation are evaluated
at imaginary frequencies, an analytic continuation to the real
frequency axis is required to compare with experimentally measured observables,
such as QP energies and spectral functions.
To this end, Thiele's reciprocal difference method~\cite{pade1975} is used.
We then apply our new implementation to predict the QP energies and spectral
functions for typical  semiconductors (Si, GaAs, SiC, and ZnO), insulators
(C, BN, MgO, and LiF),  and metals (Cu and SrVO$_3$),  and compare our
results with the conventional $GW$ implementation. For the sake of brevity,
we just show the comparison for the single-shot $GW$ calculations, i.e., $G_0W_0$, where the
one electron energies and wave functions required in $G$ and $W$  are fixed at
the DFT level. The self-consistent low-scaling $GW$ will be discussed in future publications.
To avoid confusion with the conventional $G_0W_0$, we denote our
low-scaling single-shot $GW$ as $G_0W_0r$.
It is found that the
QP energies and spectral functions predicted by $G_0W_0r$ are in good
agreement with $G_0W_0$ but with a reduced scaling in the system size
and number of $k$ points, highlighting the power of the present  method.

\section{Method } \label{sec:method}

\begin{figure}
\begin{center}
\includegraphics[width=0.49\textwidth]{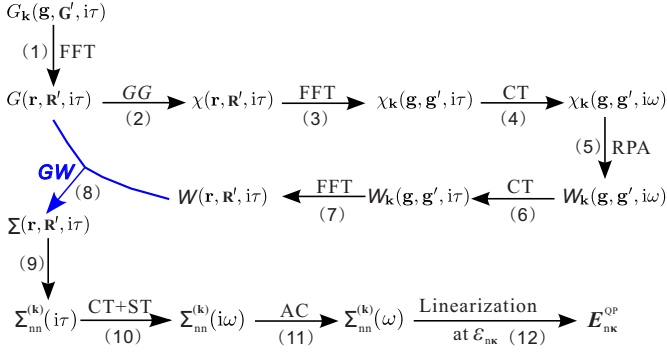}
\end{center}
\caption{Schematic work flow for the low-scaling $GW$ QP calculations
showing the necessary steps [(1)$-$(12)] to obtain the QP energies from the Green's
function $G$ via the polarizability $\chi$, screened Coulomb interaction $W$,
and self-energy $\Sigma$. FFT denote fast Fourier transformations between real and
reciprocal space. CT and ST are nonuniform cosine and sine transformations
between imaginary time and frequency.}
\label{fig:GWR}
\end{figure}

Figure \ref{fig:GWR} shows our scheme for the low-scaling $GW$ QP
calculations. The polarizability
$\chi(\rr,\RR',\ii\tau)$ is calculated via the contraction ($GG$) of the occupied and
unoccupied Green's functions within the PAW framework~\cite{KaltakPRB2014}.
The contraction is performed in real space and the necessary quantities are
obtained by fast Fourier transformations (FFT) within a supercell~\cite{KaltakPRB2014}.
Subsequently, the screened interaction $W_{\kk}(\sg,\sg',\ii\omega)$ is obtained within the RPA.
To transform the polarizability $\chi$ and
screened Coulomb interaction $W$ from imaginary time to frequency domain and vice versa,
efficient nonuniform cosine transformations (CT)~\cite{KaltakJCTC2014} are used.
The self-energy $\Sigma(\rr,\RR',\ii\tau)$  is calculated by contracting the Green's
function and $W$ within the $GW$ approximation.
The matrix elements of the self-energy in the orbital basis are evaluated within the PAW.
To transform the self-energy from the imaginary time to the frequency domain,
CT and sine transformations (ST), respectively, are used for the
even and odd parts of the self-energy.
The self-energy along the real frequency axis is obtained by an analytic continuation (AC).
Finally, the QP energies $E^{\rm QP}_\nk$ within single-shot $G_0W_0r$
are calculated by linearizing the diagonal elements of the
self-energy around the DFT one-electron eigenvalues $\epsilon_\nk$.
The subsequent subsections describe these steps in detail.

\subsection{Description of notations and definitions }
In this part, we will give the description of the notations used throughout the paper,
the definition of the Green's functions,  as well as the spatial and temporal Fourier transformations.

\subsubsection{Definitions of Green's functions}

In this paper, we have defined two types of Green's functions:
occupied and unoccupied Green's functions,
which are evaluated for the negative and positive time, respectively.
\begin{eqnarray}
\label{eq:Green_occ}
\OG(\rr,\rr', \ii\tau) =&\ \sum\limits^{\text{occ}}\limits_i \psi_i(\rr)
\psi^*_i(\rr') \ee^{-\epsilon_i\tau} \quad (\tau<0) ,  \\
\label{eq:Green_unocc}
\UG(\rr,\rr', \ii\tau)=&-\sum\limits^{\text{unocc}}\limits_a \psi_a(\rr)
\psi^*_a(\rr') \ee^{-\epsilon_a\tau} \quad (\tau>0).
\end{eqnarray}
Here the indices $i$ and $a$ label occupied and unoccupied orbitals, respectively.
$\psi_i(\rr)$ [$\psi_a(\rr)$] is the one-electron orbital with the energy of $\epsilon_i$ ($\epsilon_a$)
and the Fermi energy is set to zero.
This implies that all occupied (unoccupied) one-electron  energies $\epsilon_i$ ($\epsilon_a$)
are negative (positive), yielding exponentially decaying Green's functions $\OG$
and $\UG$.
With the definitions in Eqs. (\ref{eq:Green_occ}) and (\ref{eq:Green_unocc}),
the single-particle Green's function can be expressed as
\begin{equation}
\label{eq:single_particle_G}
G(\rr,\rr', \ii\tau) =\Theta(\tau)\UG(\rr,\rr', \ii\tau)+\Theta(-\tau)\OG(\rr,\rr', \ii\tau),
\end{equation}
where $\Theta$ is the Heaviside step function.

\subsubsection{Nonuniform imaginary time and frequency grids}
The imaginary time $\{\ii\tau_j\}_{j=1}^N$  and  frequency
$\{\ii \omega_k\}_{k=1}^N$ grids used in this work
have been determined by minimizing the discretization error
of the direct M{\o}ller-Plesset energy in the imaginary time
and frequency domain, respectively~\cite{KaltakJCTC2014}.
It has been found that the two grids are dual to each other.
That is, given $\{\ii\tau_j\}_{j=1}^N$,
the discretization error function is minimal at the
grid points $\{\ii \omega_k\}_{k=1}^N$, and vice versa~\cite{KaltakJCTC2014}.
It has also been observed that the RPA correlation energy
can be evaluated accurately with a modest number of grid points~\cite{KaltakJCTC2014}.
For instance, to achieve $\mu$eV accuracy per atom, 16 time and frequency points are usually
sufficient~\cite{KaltakJCTC2014}. In this work,
we also found that with 20 grid points, we could obtain converged
QP energies with $0.01$ eV accuracy for all the materials considered.
With a few imaginary grid points, the memory requirements are obviously much reduced.

\subsubsection{Nonuniform cosine and sine transformations}
To go from the imaginary time to imaginary frequency, and vice versa, nonuniform
discrete cosine and sine transformations have been exploited for the even and odd functions,
respectively. Specifically,  for an even function $F$ with respect to imaginary time/frequency,
such as the polarizability $\chi$  and screened Coulomb interaction $W$,
the forward Fourier transformation is given by a CT
\begin{equation}
\label{eq:CT_FFT}
F(\ii\omega_k ) =\sum\limits_{j=1}^{N} \gamma_{kj} \cos(\omega_k\tau_j) F(\ii\tau_j).
\end{equation}
Here, the imaginary time $\{\ii\tau_j\}_{j=1}^N$  and frequency $\{\ii \omega_k\}_{k=1}^N$  grids are precalculated.
The coefficients $\gamma_{kj}$ are determined in analogy to the imaginary time and frequency grids by
minimizing the error function~\cite{KaltakJCTC2014}
\begin{equation}
\label{eq:cos_error_function}
\eta^c(x,\bm{\gamma})=
\underbrace{ \ \ \ \ \ \cfrac{2x}{x^2+\omega_k^2} \ \ \ \ \ }
_{2 \int_0^\infty \mathrm{d}\tau \cos(\omega_k \tau)\mathrm{e}^{-x\tau}}
-\sum\limits_{j=1}^{N} \gamma_{kj} \cos(\omega_k\tau_j)\mathrm{e}^{-x\tau_j},
\end{equation}
for all transition energies $x \in [\epsilon_\text{min}, \epsilon_\text{max}]$
and each known frequency point $\omega_k$ separately.

Analogously, the inverse CT is given by~\cite{KaltakJCTC2014}
\begin{equation}
\label{eq:INV_CT_FFT}
F(\ii \tau_j ) = \sum\limits_{k=1}^{N}\xi_{jk} \cos(\tau_j\omega_k) F(\ii\omega_k),
\end{equation}
where the matrix $\bm{\xi\cos(\tau\omega)}$ is the inverse of the matrix $\bm{\gamma\cos(\omega\tau)}$ in Eq. (\ref{eq:CT_FFT}).

In contrast, for an odd function $F$  with respect to imaginary time/frequency,
the forward Fourier transformation is described by a ST
\begin{equation}
\label{eq:ST_FFT}
F(\ii\omega_k)=\ii \sum\limits_{j=1}^{N}\lambda_{kj} \mathrm{sin}(\omega_k \tau_j)F(\ii\tau_j).
\end{equation}
Again, $\{\ii\tau_j\}_{j=1}^N$  and $\{\ii \omega_k\}_{k=1}^N$  are precalculated and chosen to be identical to the cosine grid.
However, the coefficients $\lambda_{kj}$ are determined  by
minimizing the error function
\begin{equation}
\label{eq:sin_error_function}
\eta^s (x,\bm{\lambda})=
\underbrace{ \ \ \ \ \ \cfrac{2\omega_k}{x^2+\omega_k^2} \ \ \ \ \ }_{2 \int_0^\infty \mathrm{d}\tau \sin(\omega_k \tau)\mathrm{e}^{-x\tau}}
-\sum\limits_{j=1}^{N}\lambda_{kj} \mathrm{sin}(\omega_k \tau_j)\mathrm{e}^{-x\tau_j}.
\end{equation}
To this end,
similar strategies as discussed in Ref.~\cite{KaltakJCTC2014} are used.
The inverse ST is then obtained by
\begin{equation}
\label{eq:INV_ST_FFT}
F(\ii \tau_j ) = -\ii \sum\limits_{k=1}^{N}\zeta_{jk} \mathrm{sin}(\tau_j\omega_k) F(\ii\omega_k),
\end{equation}
where the matrix $\bm{\zeta\sin(\tau\omega)}$ is the inverse of the matrix $\bm{\lambda\sin(\omega\tau)}$ in Eq. (\ref{eq:ST_FFT}).
It should be noted that the matrices  $\bm{\gamma}$, $\bm{\xi}$, $\bm{\lambda}$, and $\bm{\zeta}$ are all precalculated
and stored after the imaginary time  $\{\ii\tau_j\}_{j=1}^N$
and frequency $\{\ii \omega_k\}_{k=1}^N$ grids are determined.

\subsubsection{Spatial fast Fourier transformation}
To transform the Green's functions from reciprocal to real space,
we employ fast discrete Fourier transformation
within a supercell~\cite{KaltakPRB2014}.  Considering the symmetry of the
Green's functions, only the irreducible stripe $G(\rr,\RR')$ needs to be
calculated in two steps~\cite{KaltakPRB2014}:
\begin{eqnarray}
\label{eq:FFT_G1}
G(\rr, \lG')=&
\sum\limits_{\sg\in\mathcal{L}_c^*} \mathrm{e}^{\ii(\kk+\sg)\rr} G_{\kk} (\sg,\sg'), \\
\label{eq:FFT_G2}
G(\rr,\RR')=&
\sum\limits_{\lG'\in\mathcal{L}_s^*} G(\rr,\lG') \mathrm{e}^{-\ii\lG'\RR'}.
\end{eqnarray}
Here, position vector $\rr$ is restricted to the unit cell ($C$), whereas $\RR$ extends over the entire supercell ($S$).
$\sg$ and $\lG$, respectively, represent the lattice vector of the reciprocal
cell ($\mathcal{L}_c^*$) and reciprocal supercell ($\mathcal{L}_s^*$). Furthermore,
$\kk$ is a $k$ point used to sample the Brillouin zone (BZ) and $\lG'=\kk+\sg'$.
The time complexity of the spatial FFT is of the order: ln$(N^2_bN_k)N^2_bN_k$
with $N_b$ and $N_k$ being the total number of considered basis vectors $\sg$
and $k$ points in the BZ, respectively~\cite{KaltakPRB2014}.

Similarly, the inverse spatial FFT is given by~\cite{KaltakPRB2014}
\begin{eqnarray}
\label{eq:INV_FFT_G1}
G(\rr,\lG')=&
\sum\limits_{\RR'\in S}
G(\rr,\RR')\mathrm{e}^{\ii \lG' \RR'},\\
\label{eq:INV_FFT_G2}
G_{\kk}(\sg,\sg')=&\sum\limits_{\rr\in C}
\mathrm{e}^{-\ii(\kk+\sg)\rr}
G(\rr,\lG'),
\end{eqnarray}
which  has the same time complexity  as the spatial FFT.

Considering that the polarizability $\chi$ has the same spatial symmetry as the Green's functions,
the above mentioned spatial and inverse spatial FFT applies to the
polarizability $\chi$ as well.

\subsection{Calculation of the polarizability $\chi(\rr,\RR',i\tau)$ within the PAW}
In this section, we discuss the steps (1--2) in Fig. \ref{fig:GWR} and
derive a suitable expression for the polarizability in real space
$\chi(\rr,\RR',\ii\tau)$ within the framework of the PAW method.

It is known that the evaluation of the polarizability
in reciprocal space and real frequency results in an unfavorable scaling.
However, the polarizability is simply multiplicative,
when evaluated in the real space and imaginary time domain~\cite{Godby1995,Godby2000}
\begin{equation}\label{eq:chi_GG}
\chi(\rr,\RR',\ii\tau)=G(\rr,\RR',\ii\tau)G^*(\rr,\RR',-\ii\tau).
\end{equation}
For simplicity, we restrict our considerations to positive imaginary times
$\tau>0$ in the following, since the expressions for $\tau<0$ are obtained by
exchanging $\UG \leftrightarrow \OG$.

Inserting expression (\ref{eq:single_particle_G}) for the Green's function into
(\ref{eq:chi_GG}) and using the explicit representations
in Eqs.~(\ref{eq:Green_occ}--\ref{eq:Green_unocc}) yields for $\tau>0$,
\begin{equation}\label{eq:chi_rr}
\chi(\rr,\RR',\ii\tau)=-\sum^{\text{unocc}}_a \psi_a(\rr)\psi_a^*(\RR')
\ee^{-\epsilon_a\tau}
\sum^{\text{occ}}_i\psi_i(\RR')\psi_i^*(\rr)\ee^{\epsilon_i\tau}.
\end{equation}
Fourier transforming this expression using Eqs.
(\ref{eq:INV_FFT_G1}--\ref{eq:INV_FFT_G2}) gives the polarizability in reciprocal
space for $\tau>0$,
\begin{equation}\label{eq:CHI_g_it}
\begin{split}
\chi_\kk(\sg,\sg',\ii\tau)
=
-\sum^{\text{occ}}_i \sum^{\text{unocc}}_a \xi_{ia}(\ii\tau)
& \bra{\psi_i} e^{-\ii (\kk+\sg) \rr} \ket{\psi_a} \\
\times &
\bra{\psi_a} e^{\ii (\kk+\sg') \rr'} \ket{\psi_i},
\end{split}
\end{equation}
where $\xi_{ia}(\tau)=\ee^{-(\epsilon_a-\epsilon_i)\tau}$ describes the time
dependence.

However, within the PAW method~\cite{PAW1994,Kresse1999}, this expression is more
involved, because the all-electron orbitals $\psi_\nk$ are related to the
corresponding pseudo orbitals $\tilde\psi_\nk$ by a linear transformation
\begin{equation} \label{eq:PAW_psi}
\ket{\psi_\nk }=\ket{\tilde\psi_\nk }+\sum_\mu \left(\ket{\phi_\mu}-\ket{\tilde\phi_\mu}\right)
\bra{\tilde{p}_\mu}\tilde\psi_\nk\rangle.
\end{equation}
The  pseudo orbitals $\tilde\psi_\nk$ are the variational quantity of the PAW
method and are expanded in plane waves, whereas $\phi_\mu$  and $\tilde\phi_\mu$ are all-electron
and pseudo partial waves, respectively and $\tilde{p}_\mu$ are projectors, which are
dual to the $\tilde\phi_\mu$ within the augmentation sphere.
The index $\mu=(\mathbf{R}_\mu,n_\mu,l_\mu,m_\mu)$ is an
abbreviation for the atomic site $\mathbf{R}_\mu$ and the energy quantum number
$n_\mu$ and angular momentum numbers $(l_\mu,m_\mu)$ that characterize the
solution of the Schr\"odinger equation for a reference atom.

Inserting Eq. (\ref{eq:PAW_psi}) into Eq. (\ref{eq:chi_rr})
yields four terms for $\tau>0$,
\begin{equation}\label{eq:CHI_rR_T1_final}
\begin{split}
\chi^{(1)}(\rr,\RR',\ii\tau)
=-
\sum^{\text{unocc}}_a \ppsi_a(\rr)\ppsi^*_a(\RR')\ee^{-\epsilon_a\tau}
\sum^{\text{occ}}_i\ppsi_i(\RR') \ppsi^*_i(\rr) \ee^{\epsilon_i\tau},
\end{split}
\end{equation}

\begin{equation}\label{eq:CHI_rR_T2_final}
\begin{split}
\chi^{(2)}(\rr,\RR',\ii\tau)
=-
\sum\limits_{\mu\nu} Q_{\mu\nu}(\rr)
&  \sum^{\text{unocc}}_a \sandw{\pp_\nu}{\ppsi_{a}}\ppsi^*_a(\RR') \ee^{-\epsilon_a\tau} \\
\times & \sum^{\text{occ}}_i  \ppsi_i(\RR')\sandw{\ppsi_i}{\pp_\mu}\ee^{\epsilon_i\tau},
\end{split}
\end{equation}

\begin{equation}\label{eq:CHI_rR_T3_final}
\begin{split}
\chi^{(3)}(\rr,\RR',\ii\tau)
=-
\sum\limits_{\alpha\beta}Q_{\alpha\beta}(\RR')
&\sum^{\text{unocc}}_a \ppsi_a(\rr) \sandw{\ppsi_a}{\pp_\alpha}\ee^{-\epsilon_a\tau} \\
\times & \sum^{\text{occ}}_i\sandw{\pp_\beta}{\ppsi_i} \ppsi^*_i(\rr) \ee^{\epsilon_i\tau},
\end{split}
\end{equation}

\begin{equation}\label{eq:CHI_rR_T4_final}
\begin{split}
\chi^{(4)}(\rr,\RR',\ii\tau)
=-
& \sum\limits_{\mu\nu\alpha\beta}Q_{\mu\nu}(\rr)Q_{\alpha\beta}(\RR') \\
\times &
\sum^{\text{unocc}}_a  \sandw{\pp_\nu}{\ppsi_a}\sandw{\ppsi_a}{\pp_\alpha}
  \ee^{-\epsilon_a\tau}  \\
\times &
 \sum^{\text{occ}}_i
\sandw{\pp_\beta}{\ppsi_i}\sandw{\ppsi_i}{\pp_\mu}\ee^{\epsilon_i\tau},
\end{split}
\end{equation}
where the auxiliary function $Q_{\alpha\beta}(\rr)$ is defined as
\begin{eqnarray}
\label{eq:pawcharge}
Q_{\alpha\beta}(\rr)=\phi^*_\alpha(\rr)\phi_\beta(\rr)-\tilde\phi^*_\alpha(\rr)\tilde\phi_\beta(\rr),
\end{eqnarray}
and describes the difference between the charge density of  all-electron
and pseudo partial waves.
In practice, one needs further approximations for
$Q_{\alpha\beta}(\rr)$, since normally this function is oscillatory
within the augmentation sphere. In the present implementation,
the function is expanded in an orthogonal set of functions,
and the rapid spatial oscillations are neglected
beyond a certain plane-wave energy cutoff~\cite{Kresse1999}.

According to the definitions of the Green's functions in
Eqs. (\ref{eq:Green_occ}) and (\ref{eq:Green_unocc}) and four expressions in Eqs.
(\ref{eq:CHI_rR_T1_final}) -- (\ref{eq:CHI_rR_T4_final}), we define here four auxiliary
functions for unoccupied Green's functions:
\begin{eqnarray}
\label{eq:G1_unocc_r}
\overline{G}^{(1)}(\rr,\RR', i\tau)&=&-\sum\limits^{\rm unocc}_a\ppsi_a(\rr)\ppsi^*_a(\RR') \ee^{-\epsilon_a\tau}, \\
\label{eq:G2_unocc_r}
\overline{G}^{(2)}(\nu,\RR', i\tau)&=&-\sum\limits^{\rm unocc}_a\sandw{\pp_\nu}{\tpsi_a}\ppsi^*_a(\RR') \ee^{-\epsilon_a\tau}, \\
\label{eq:G3_unocc_r}
\overline{G}^{(3)}(\rr,\alpha, i\tau)&=&-\sum\limits^{\rm unocc}_a \ppsi_a(\rr) \sandw{\tpsi_a}{\pp_\alpha}\ee^{-\epsilon_a\tau},  \\
\label{eq:G4_unocc_r}
\overline{G}^{(4)}(\nu,\alpha, i\tau)&=&-\sum\limits^{\rm unocc}_a\sandw{\pp_\nu}{\tpsi_a} \sandw{\tpsi_a}{\pp_\alpha} \ee^{-\epsilon_a\tau},
\end{eqnarray}
and four auxiliary functions for occupied  Green's functions:
\begin{eqnarray}
\label{eq:G1_occ_r}
\underline{G}^{*(1)}(\rr,\RR', -i\tau)&= &\sum\limits^{\rm occ}_i \ppsi_i(\RR')\ppsi^*_i(\rr)\ee^{\epsilon_i\tau}, \\
\label{eq:G2_occ_r}
\underline{G}^{*(2)}(\mu,\RR' , -i\tau)&=& \sum\limits^{\rm occ}_i \ppsi_i(\RR')
\sandw{\ppsi_i}{\pp_\mu}\ee^{\epsilon_i\tau}, \\
\label{eq:G3_occ_r}
\underline{G}^{*(3)}(\rr,\beta, -i\tau)&=& \sum\limits^{\rm occ}_i
\sandw{\pp_\beta}{\ppsi_i}\ppsi^*_i(\rr)\ee^{\epsilon_i\tau},  \\
\label{eq:G4_occ_r}
\underline{G}^{*(4)}(\mu,\beta, -i\tau)&=& \sum\limits^{\rm occ}_i
\sandw{\pp_\beta}{\tpsi_i} \sandw{\tpsi_i}{\pp_\mu} \ee^{\epsilon_i\tau}.
\end{eqnarray}
It is easy to prove that
\begin{eqnarray}
\label{eq:G21_occ_relation}
{G}^{(2)}(\nu,\RR', i\tau)&=& \sum\limits_{\rr\in C}  \sandw{\pp_\nu}{\rr} {G}^{(1)}(\rr,\RR', i\tau),\\
\label{eq:G31_occ_relation}
{G}^{(3)}(\rr,\alpha, i\tau)&=&  \sum\limits_{\RR'\in S} {G}^{(1)}(\rr,\RR', i\tau) \sandw{\RR'}{\pp_\alpha}, \\
\label{eq:G41_occ_relation}
{G}^{(4)}(\nu,\alpha, i\tau)&=&\sum\limits_{\rr\in C} \sum\limits_{\RR'\in S} \sandw{\pp_\nu}{\rr} {G}^{(1)}(\rr,\RR', i\tau)
\sandw{\RR'}{\pp_\alpha},
\end{eqnarray}
holds for both auxiliary unoccupied and  occupied Green's functions.

With the definitions in Eqs. (\ref{eq:G1_unocc_r})--(\ref{eq:G4_occ_r}),
we obtain the central expression for the polarizability $\chi(\rr,\RR',i\tau)$ at $\tau>0$
within the PAW framework as follows:
\begin{equation}\label{eq:CHI_rR_all_final}
\begin{split}
\chi(\rr,\RR',\ii\tau)
=&
\overline{G}^{(1)}(\rr,\RR', i\tau)
\underline{G}^{*(1)}(\rr,\RR', -i\tau) \\
+&
\sum\limits_{\mu\nu} Q_{\mu\nu}(\rr)
\overline{G}^{(2)}(\nu,\RR', i\tau)
\underline{G}^{*(2)}(\mu,\RR', -i\tau) \\
+&
\sum\limits_{\alpha\beta}Q_{\alpha\beta}(\RR')
\overline{G}^{(3)}(\rr,\alpha, i\tau)
\underline{G}^{*(3)}(\rr,\beta, -i\tau)  \\
+&
\sum\limits_{\mu\nu\alpha\beta}Q_{\mu\nu}(\rr)Q_{\alpha\beta}(\RR')
\overline{G}^{(4)}(\nu,\alpha, i\tau)
\underline{G}^{*(4)}(\mu,\beta, -i\tau).
\end{split}
\end{equation}
Here, the atomic positions $\RR_\mu$, $\RR_\nu$  are restricted to the unit cell $C$,
while $\RR_\alpha$, $\RR_\beta$ take values within the supercell $S$.
Note that the polarizability for $\tau<0$ is recovered from Eq. (\ref{eq:CHI_rR_all_final})
 by exchanging $\UG \leftrightarrow \OG$.

In practice, we do not store the auxiliary Green's functions in
Eqs. (\ref{eq:G1_unocc_r})--(\ref{eq:G3_unocc_r}) and
 (\ref{eq:G1_occ_r})--(\ref{eq:G3_occ_r}) directly using the real-space grids,
since this would demand considerable storage due to the large
number of real-space grid points.
 Instead, we evaluate them in the reciprocal space using a plane wave
representation first, and successively Fourier transform the functions to real space
whenever required. Since the number of plane-wave coefficients is
at least twice but often up to 16 times smaller than the number of real-space grid points,
the storage demand is dramatically reduced.
Fourier transforming Eqs. (\ref{eq:G1_unocc_r})--(\ref{eq:G3_unocc_r})  to
the reciprocal space yields another three auxiliary unoccupied Green's functions:
\begin{eqnarray}
\label{eq:G1_unocc_k}
\overline{G}^{(1)}_\kk(\sg,\lG', i\tau)&= &-\sum\limits^{\rm unocc}_a \sandw{\sg}{\tpsi_a}\sandw{\tpsi_a}{\lG'}\ee^{-\epsilon_a\tau}, \\
\label{eq:G2_unocc_k}
\overline{G}^{(2)}_\kk(\nu,\lG', i\tau)&=&-\sum\limits^{\rm unocc}_a\sandw{\pp_\nu}{\tpsi_a}\sandw{\ppsi_a}{\lG'} \ee^{-\epsilon_a\tau}, \\
\label{eq:G3_unocc_k}
\overline{G}^{(3)}_\kk(\sg,\alpha, i\tau)&=&-\sum\limits^{\rm unocc}_a \sandw{\sg}{\ppsi_a}\sandw{\tpsi_a}{\pp_\alpha}\ee^{-\epsilon_a\tau}.
\end{eqnarray}
Analogously, Fourier transforming Eqs. (\ref{eq:G1_occ_r})--(\ref{eq:G3_occ_r})  to
the reciprocal space yields another three auxiliary occupied Green's functions~\cite{note}:
\begin{eqnarray}
\label{eq:G1_occ_k}
\underline{G}^{*(1)}_\kk(\sg,\lG', -i\tau)&=&\sum\limits^{\rm occ}_i
\sandw{\lG'}{\tpsi_i}\sandw{\tpsi_i}{\sg}\ee^{\epsilon_i\tau}, \\
\label{eq:G2_occ_k}
\underline{G}^{*(2)}_\kk(\mu,\lG', -i\tau)&= &\sum\limits^{\rm occ}_i
 \sandw{\lG'}{\ppsi_i} \sandw{\ppsi_i}{\pp_\mu}\ee^{\epsilon_i\tau}, \\
\label{eq:G3_occ_k}
\underline{G}^{*(3)}_\kk(\sg,\beta, -i\tau)&= & \sum\limits^{\rm occ}_i
\sandw{\pp_\beta}{\ppsi_i}\sandw{\ppsi_i}{\sg}\ee^{\epsilon_i\tau},
\end{eqnarray}
where the notation
\begin{eqnarray}\label{eq:notation_bra_ket}
\sandw{\sg}{\tpsi} &=& \sum_{\rr \in C} \ee^{-\ii(\kk+\sg)\rr} \tpsi(\rr), \\
\sandw{\tpsi}{\lG'} &=& \sum_{\RR' \in S}  \tpsi^*(\RR') \ee^{\ii\lG'\RR'},
\end{eqnarray}
is used. The computational complexity for evaluating both $\OG^{(j)}$ and $\UG^{(j)}$ is
of the order: $NN_k N^3_b$ showing a roughly cubic scaling in the system size
($\approx$$N_b$) and linear scaling in the number of $k$ points $N_k$ and imaginary grid points
$N$.

We point out that
the widely used conventional $GW$ implementation~\cite{Shishkin2006},
where the polarizability is directly evaluated in the reciprocal space and real
frequency domain, shows an unfavorable scaling that is
quartic in the system size and quadratic in the number of $k$ points.
This scaling is acceptable or even beneficial for small systems, but prohibitive as the system size becomes
larger. In contrast, in our new $GW$ implementation,
the computational cost in calculating the polarizability reduces to a scaling that is nearly
cubic in the system size and linear in the number of $k$ points. This definitely increases
the efficiency of $GW$ calculations for large systems.
This is also true for the original space-time implementation of Godby \emph{et al.}~\cite{Godby1995,Godby2000},
but we emphasize that our implementation has the following advantages:
(i) The Green's functions are stored in a plane wave representation at
a few optimized imaginary time/frequency grid points, which
dramatically reduces the memory requirement.
(ii) It is implemented within the PAW method.
(iii) Discrete CT and ST transformations and spatial FFT
are used and the implementation is highly parallelized.
(iv) Although similar strategies were used in Ref.~\cite{Godby2000},
an auxiliary supercell Green's function was defined
without the Bloch phase factors $\ee^{\ii\kk(\rr-\rr')}$.
The present method is applicable to all electron Hamiltonians, whereas
the augmentation terms cannot be straightforwardly implemented following
Ref.~\cite{Godby2000}.

\subsection{Calculation of the correlated screened Coulomb interaction $\tilde W(\rr,\RR',\ii\omega)$ }

Now we describe the evaluation of the dynamical correlated  screened Coulomb interaction $\tilde W(\rr,\RR',\ii\omega)$,
which corresponds to the steps (3--7) in Fig. \ref{fig:GWR}.

Once the polarizability $\chi(\rr,\RR',\ii\tau)$ has been
calculated, one has to Fourier transform it to the reciprocal space and imaginary frequency domain
where the screened Coulomb interaction is much more comfortable to be calculated. The calculation of
$\tilde W(\rr,\RR',\ii\omega)$  involves five steps:

(i) $\chi_{\kk}(\sg,\sg',\ii\tau)$ is determined by an inverse spatial FFT of $\chi(\rr,\RR',\ii\tau)$ in two steps:
\begin{eqnarray}
\label{eq:FFTX1}
\chi(\rr,\lG',\ii\tau)=&
\sum\limits_{\RR'\in S}
\chi(\rr,\RR',\ii\tau)\mathrm{e}^{\ii \lG' \RR'},\\
\label{eq:FFTX2}
\chi_{\kk}(\sg,\sg',\ii\tau)=&\sum\limits_{\rr\in C}
\mathrm{e}^{-\ii(\kk+\sg)\rr}
\chi(\rr,\lG',\ii\tau).
\end{eqnarray}
Actually, in our implementation the polarizability $\chi(\rr,\RR',\ii\tau)$ is never stored.
Instead, once $\chi(\rr,\RR',\ii\tau)$ is known for a specific $\rr$ and all $\RR'$,  Eq. (\ref{eq:FFTX1})
is used to Fourier transform the second index to the reciprocal space where
the reciprocal wave vectors are restricted to a cutoff sphere, and $\chi(\rr,\lG',\ii\tau)$
is then stored. The second FFT in Eq. (\ref{eq:FFTX2}) cannot be performed until
$\chi(\rr,\lG',\ii\tau)$ for all $\rr$ has been calculated.

(ii) $\chi_{\kk}(\sg,\sg',\ii \omega)$ is computed by a CT of $\chi_{\kk}(\sg,\sg',\ii\tau)$,
\begin{equation}
\label{eq:CTX}
\chi_{\kk}(\sg,\sg',\ii \omega_k ) =
\sum\limits_{j=1}^{N} \gamma_{kj} \cos(\omega_k\tau_j) \chi_{\kk}(\sg,\sg',\ii\tau_j).
\end{equation}

(iii) The full screened Coulomb interaction $W_\kk(\sg,\sg',\ii\omega)$ is evaluated
by multiplying the bare Coulomb kernel with the inverse dielectric matrix,
\begin{equation}
\label{eq:W_tot}
W_\kk(\sg,\sg',\ii\omega)=v_\kk(\sg,\sg')
\varepsilon_\kk^{-1}(\sg,\sg',\ii\omega),
\end{equation}
where the symmetric bare Coulomb kernel $v_\kk(\sg,\sg')$ is
\begin{equation} \label{eq:bareV}
v_\kk(\sg,\sg')= \cfrac{4\pi e^2}{|\kk+\sg||\kk+\sg'|}.
\end{equation}
The symmetric dielectric matrix is calculated within the RPA as
\begin{equation}
\label{eq:die}
\varepsilon_\kk(\sg,\sg',\ii\omega)=\delta_{\sg,\sg'}
-v_\kk(\sg,\sg') \chi_\kk(\sg,\sg',\ii\omega).
\end{equation}

To make the integral over the imaginary frequency well-defined,
we further define the correlated screened Coulomb interaction
\begin{equation} \label{eq:W_correlated}
\tilde W_\kk(\sg,\sg',\ii\omega)=W_\kk(\sg,\sg',\ii\omega)-v_\kk(\sg,\sg').
\end{equation}

(iv)  $\tilde W_\kk(\sg,\sg',\ii\tau)$ is determined by an inverse CT of $\tilde W_\kk(\sg,\sg',\ii\omega)$,
\begin{equation}
\label{CT_W}
\tilde W_\kk(\sg,\sg',\ii \tau_j ) =
\sum\limits_{k=1}^{N}\xi_{jk} \cos(\tau_j\omega_k) \tilde W_\kk(\sg,\sg',\ii\omega_k).
\end{equation}

(iv) Finally, $\tilde W(\rr,\RR',\ii\tau)$  is calculated by a spatial FFT in two steps:
\begin{eqnarray}
\label{eq:FFTW}
\tilde W(\rr, \lG' ,\ii\tau)=&
\sum\limits_{\sg\in\mathcal{L}_c^*}
\mathrm{e}^{\ii(\kk+\sg)\rr} \tilde W_{\kk} (\sg,\sg',\ii\tau),  \\
\tilde W(\rr,\RR',\ii\tau)=&\sum\limits_{\lG'\in\mathcal{L}_s^*}
\tilde W(\rr,\lG',\ii\tau) \mathrm{e}^{-\ii\lG'\RR'}.
\end{eqnarray}

\subsection{Calculation of the self-energy}
In this section, we give a detailed description how the matrix elements of the self-energy in the
orbital basis along the imaginary frequency axis are evaluated.
This corresponds to the steps (8--10) in Fig. \ref{fig:GWR}.

\subsubsection{Evaluation of the self-energy $\Sigma(\rr,\RR',\ii\tau)$  within
the $GWA$ }
Within the $GW$ approximation, the self-energy in the reciprocal space and real frequency domain
 is evaluated by a convolution of the Green's function  and screened Coulomb
interaction and is used in the conventional $GW$ implementation~\cite{Shishkin2006}.
However, to obtain converged self-energies, a reasonable
number of the real frequency points ($\sim$50 or more) is
required to evaluate the convolution integral, thus increasing the computational cost.
In contrast, when the self-energy is evaluated in real space
and time, it is simply multiplicative~\cite{Godby1995}
\begin{equation}
\Sigma(\rr,\RR',\ii\tau)=-G(\rr,\RR',\ii\tau)W(\rr,\RR',\ii\tau).
\end{equation}
In addition, only a few imaginary time points are required due to the smooth behavior of the
Green's functions and screened Coulomb interaction along the imaginary axis.

\subsubsection{Evaluation of $\tilde \Sigma^{(\kk)}_{nn}(\ii\tau)$ within the PAW}

In the following, we evaluate the matrix elements of the self-energy in the
orbital basis within the PAW framework. We focus only on the frequency/time-dependent
correlation contribution $-\bra{\psi_\nk}G \tilde W\ket{\psi_\nk}$,
since the bare exchange part $-\bra{\psi_\nk}G v_x\ket{\psi_\nk}$ within the
PAW has already been discussed elsewhere~\cite{HSE2005}.
Furthermore, we define {\lq\lq}occupied{\rq\rq} $\tilde \OS$ and {\lq\lq}unoccupied{\rq\rq} $\tilde \US$  correlated self-energies,
i.e.,  the self-energies evaluated at negative and positive time, respectively, analogous to the Green's functions.

Here we concentrate on the occupied self-energy  $\tilde \OS$ only. The
evaluation of the matrix elements of the unoccupied self-energy  $\tilde \US$
is done by replacing $\underline G$ with $\overline G$.
Within the PAW, the diagonal matrix elements of the
occupied self-energy ($\tau<0$) can be calculated as
\begin{equation}\label{eq:Sigma_occ}
\begin{split}
\tilde{\OS}^{(\kk)}_{\text{nn}}(\ii\tau)
&=\bra{\psi_\nk}\tilde{\OS}(\ii\tau) \ket{\psi_\nk} = -\bra{\psi_\nk} \OG(\ii\tau) \tilde W(\ii\tau)\ket{\psi_\nk} \\
&=-\sum\limits_{\rr \in C} \sum\limits_{\RR' \in S} \bra{\tilde \psi_\nk}
\left \{ \ket{\rr}\bra{\rr}+\sum_{\mu\nu} Q_{\mu\nu}(\rr) \ket{\pp_\mu}\bra{\pp_\nu} \right \}
 \OG (\ii\tau)  \\
& \times \tilde W(\ii\tau)
 \left \{\ket{\RR'}\bra{\RR'}+\sum_{\alpha\beta} Q_{\alpha\beta}(\RR') \ket{\pp_\alpha}\bra{\pp_\beta} \right \}  \ket{\tilde \psi_\nk}.
\end{split}
\end{equation}
Here $\ket{\rr}\bra{\rr}+\sum_{\mu\nu} Q_{\mu\nu}(\rr) \ket{\pp_\mu}\bra{\pp_\nu}$ is
the density operator within the PAW at the position $\rr$~\cite{PAW1994,Kresse1999}. The one-center term
$\sum_{\mu\nu} Q_{\mu\nu}(\rr) \ket{\pp_\mu}\bra{\pp_\nu}$ arises from the additive augmentation of the PAW.

The calculation is performed in two steps, starting with the contraction $\tilde{\Sigma}=-G\tilde{W}$
in real space and imaginary time.
In analogy to the four auxiliary components of the Green's function, we obtain four quantities that store the
self-energy,
\begin{eqnarray}
\label{eq:Sigma_rr}
\tilde{\OS}^{(1)}(\rr,\RR', \ii\tau) &=&-\OG^{(1)}(\rr,\RR', \ii\tau) \tilde{W}(\rr,\RR', \ii\tau), \\
\label{eq:Sigma_ra}
\tilde{\OS}^{(2)}(\mu,\RR', \ii\tau) &=&-\sum\limits_{\nu} D^{(2)}(\mu \nu,\RR', \ii\tau) \OG^{(2)} (\nu,\RR', \ii\tau),\\
\label{eq:Sigma_ar}
\tilde{\OS}^{(3)}(\rr,\beta, \ii\tau) &=&-\sum\limits_{\alpha} \OG^{(3)}(\rr,\alpha, \ii\tau) D^{(3)}(\rr,\alpha\beta, \ii\tau), \\
\label{eq:Sigma_aa}
\tilde{\OS}^{(4)}(\mu,\beta, \ii\tau)&=&-\sum\limits_{\nu\alpha} \OG^{(4)}(\nu,\alpha, \ii\tau) D^{(4)}(\mu \nu,\alpha\beta, \ii\tau),
\end{eqnarray}
where the auxiliary quantities $D^{(2)}$, $D^{(3)}$ and $D^{(4)}$ are defined as
\begin{eqnarray}
\label{eq:D2}
D^{(2)}(\mu \nu,\RR', \ii\tau) &=& \sum\limits_{\rr \in C} Q_{\mu\nu}(\rr) \tilde{W}(\rr,\RR', \ii\tau) , \\
D^{(3)}(\rr,\alpha\beta, \ii\tau) &=& \sum\limits_{\RR' \in S} \tilde{W}(\rr,\RR', \ii\tau) Q_{\alpha\beta}(\RR'), \\
D^{(4)}(\mu \nu,\alpha\beta, \ii\tau) &=& \sum\limits_{\rr \in C} \sum\limits_{\RR' \in S} Q_{\mu\nu}(\rr) \tilde{W}(\rr,\RR', \ii\tau) Q_{\alpha\beta}(\RR').
\end{eqnarray}
Again, the Green's functions and the screened interaction are stored in reciprocal space and Fourier transformed
to the real space on the fly, whenever they are required.
To reduce the memory requirements, the self-energy is also stored in reciprocal space.
In the second step, the matrix elements of the self-energy are then obtained as
\begin{equation}
\label{eq:Sigma_occ_reduce}
\begin{split}
\tilde{\OS}^{(\kk)}_{\text{nn}}(\ii\tau)
  &=\sum\limits_{\rr \in C} \sum\limits_{\RR' \in S} \tpsi^*_\nk(\rr) \tilde{\OS}^{(1)}(\rr,\RR', \ii\tau) \tpsi_\nk(\RR') \\
   &+\sum\limits_{\mu} \sum\limits_{\RR' \in S} \sandw{\tpsi_\nk}{\pp_\mu} \tilde{\OS}^{(2)}(\mu,\RR', \ii\tau) \tpsi_\nk(\RR') \\
   &+\sum\limits_{\rr \in C}\sum\limits_{\beta} \tpsi^*_\nk(\rr) \tilde{\OS}^{(3)}(\rr,\beta, \ii\tau) \sandw{\pp_\beta}{\tpsi_\nk} \\
   &+\sum\limits_{\mu}\sum\limits_{\beta}\sandw{\tpsi_\nk}{\pp_\mu}  \tilde{\OS}^{(4)}(\mu,\beta, \ii\tau) \sandw{\pp_\beta}{\tpsi_\nk}.\\
\end{split}
\end{equation}

One point that should be mentioned here is that in the present implementation,
the core-valence exchange-correlation interaction is treated in the same way
as in the conventional $GW$ implementation, that is, the Hartree-Fock approximation is used.
This is found to be more reliable than LDA since the $GW$ self-energy approaches
the bare Fock exchange operator in the short wavelength limit~\cite{Shishkin2006}.

\subsubsection{Evaluation of  $\tilde \Sigma^{(\kk)}_{nn}(\ii\omega)$ by \text{CT+ST}}
After the matrix elements of the self-energy along the imaginary time have been obtained,
one needs to Fourier transform them to the imaginary frequency domain to calculate
the QP energies. However, the self-energy (like the Green's function) is neither
an even nor an odd function in imaginary time/frequency. Hence,
we split the Green's functions into even and  odd  parts,
\begin{equation}\label{eq:Green_split}
G(\ii\tau)=
  \cfrac{1}{2}\left[ G(\ii\tau)+G(-\ii\tau) \right]
+\cfrac{1}{2}\left[ G(\ii\tau)-G(-\ii\tau) \right].
\end{equation}

Then, the self-energy along the imaginary frequency is given by
the temporal Fourier transformation
\begin{equation}\label{eq:FFT_Sigma}
\begin{split}
\tilde \Sigma(\ii\omega)
= & -\int\limits_{-\infty}^{\infty}\mathrm{d}\tau G(\ii\tau) \tilde  W(\ii\tau)\mathrm{e}^{\ii\omega\tau} \\
= &
2\int\limits_{0}^{\infty}\mathrm{d}\tau \tilde \Sigma^c(\ii\tau)\cos(\omega\tau) +
2\ii\int\limits_{0}^{\infty}\mathrm{d}\tau \tilde \Sigma^s(\ii \tau)\sin(\omega\tau),
\end{split}
\end{equation}
where the cosine $\tilde \Sigma^c$ and sine $\tilde \Sigma^s$ part read
\begin{align}\label{eq:cos_sin_sigma}
\tilde \Sigma^c(\ii\tau)=&
- \cfrac{1}{2} \left[ \UG(\ii\tau)+\OG(-\ii\tau) \right] \tilde  W(\ii \tau), \\
\tilde \Sigma^s(\ii\tau)=&
- \cfrac{1}{2} \left[\UG(\ii\tau)-\OG(-\ii\tau) \right] \tilde W(\ii \tau).
\end{align}
Therefore, the corresponding diagonal matrix elements are given by
\begin{align}
\label{eq:cos_sin_sigma_matrix1}
\tilde \Sigma^{c(\kk)}_{\text{nn}}(\ii\tau) =&
 \cfrac{1}{2} \left[ \overline{\tilde \Sigma}^{(\kk)}_{\text{nn}}(\ii\tau) +\underline{\tilde \Sigma}^{(\kk)}_{\text{nn}}(-\ii\tau) \right], \\
\label{eq:cos_sin_sigma_matrix2}
 \tilde \Sigma^{s(\kk)}_{\text{nn}}(\ii\tau)=&
  \cfrac{1}{2} \left[ \overline{\tilde \Sigma}^{(\kk)}_{\text{nn}}(\ii\tau) -\underline{\tilde \Sigma}^{(\kk)}_{\text{nn}}(-\ii\tau) \right].
\end{align}

Finally, the diagonal matrix elements of the correlated self-energy along the
imaginary frequency axis are evaluated as
\begin{equation}
\label{eq:sigma_r_w}
\tilde \Sigma^{(\kk)}_{\text{nn}}(\ii\omega)=\tilde \Sigma^{c(\kk)}_{\text{nn}}(\ii\omega)+\tilde \Sigma^{s(\kk)}_{\text{nn}}(\ii\omega),
\end{equation}
where $\tilde \Sigma^{c(\kk)}_{\text{nn}}(\ii\omega)$ and $\tilde \Sigma^{s(\kk)}_{\text{nn}}(\ii\omega)$,
respectively, are determined by discrete CT and ST:
\begin{eqnarray}
\label{eq:cos_sin_sigma1}
\tilde \Sigma^{c(\kk)}_{\text{nn}}(\ii\omega_k)=&
\sum\limits_{j=1}^{N}\gamma_{kj} \mathrm{cos}(\omega_k \tau_j) \tilde \Sigma^{c(\kk)}_{\text{nn}}(\ii\tau_j), \\
\label{eq:cos_sin_sigma2}
\tilde \Sigma^{s(\kk)}_{\text{nn}}(\ii\omega_k)=&
\ii \sum\limits_{j=1}^{N}\lambda_{kj} \mathrm{sin}(\omega_k \tau_j)  \tilde \Sigma^{s(\kk)}_{\text{nn}}(\ii\tau_j).
\end{eqnarray}

\subsection{Calculation of QP energies and spectral functions}
In this section, we describe the calculation of QP energies and spectral functions,
which corresponds to the last two steps in Fig. \ref{fig:GWR}.

\subsubsection{Analytic continuation}\label{sec:AC}
In our present implementation, the self-energy and Green's function are
 calculated in the imaginary frequency domain.
However, the experimental observables of interest, such as QP energies and
spectral functions, are obviously measured
all along the real frequency axis. This implies that an analytic continuation from the imaginary
to the real frequency domain has to be performed. Given that our self-energy is exact in the sense
that there are no stochastic noises [unlike the Green's functions $G(\ii\tau)$ obtained from
quantum Monte-Carlo (QMC) simulations], here we utilize the $N$-point Pad\'{e}
approximant and  employ Thiele's reciprocal difference method~\cite{pade1975}.
\begin{eqnarray}
\label{pade1}
P_N(z)=\cfrac{a_1}{1+} \  \cfrac{a_2(z-z_1)}{1+} \ \cdot\cdot\cdot\ \cfrac{a_N(z-z_{N-1})}{1+(z-z_N)g_{N+1}(z)}.
\end{eqnarray}
Here, the complex coefficients $a_n$  are obtained by the following recursion relations:
\begin{eqnarray}
\label{pade2}
a_n=g_n(z_n), \ \     g_1(z_n)=f_n,  \ \     n=1, \cdot\cdot\cdot, N \\
g_n(z)=\cfrac{g_{n-1}(z_{n-1})-g_{n-1}(z)}{(z-z_{n-1})g_{n-1}(z)}, \ \  n\geq2.
\end{eqnarray}
It is straight forward to prove~\cite{pade1975} that $P_N(z_j)=f_j$ holds for the known point-value
pairs $\{z_i,f_j\}_{j=1}^N$ of the function $f(z)$
(the diagonal elements of the self-energy in the $G_0W_0r$ case).
It should be emphasized that Thiele's reciprocal difference method is fairly stable,
whereas a naive computation of Pad\'{e} coefficients usually yields
numerical instabilities.
Thiele's reciprocal difference method has been successfully applied to the analytic
continuation of dynamic response functions~\cite{Lee1996}. In the following we show that
this method can be used for the accurate prediction of
$GW$ QP energies and spectral functions as well (see Section \ref{sec:results}).

\subsubsection{Evaluation of $E^{QP}_\nk$  and $A_\nk(\omega)$}
After the diagonal elements of the self-energy along the real frequency axis $\Sigma^{(\kk)}_{\text{nn}}(\omega)$
including contributions from the core-valence exchange-correlation,
bare exchange and dynamical interactions [hereafter we denote it as $\Sigma_\nk(\omega)$]
have been obtained by the analytic continuation,
the QP energies are evaluated as in conventional $GW$ implementations.
This means, for the single-shot $GW$ calculations, the QP energies are
calculated to first order, by linearizing the self-energy around the DFT
single particle eigenvalues $\epsilon_\nk$:
\begin{equation}
\label{eq: QP}
\Sigma_\nk(E^{\rm QP}_\nk)=\Sigma_\nk(\epsilon_\nk) +
\left.\cfrac{\partial\Sigma_\nk(\omega)}{\partial\omega}
\right|_{\omega=\epsilon_\nk}(E^{\rm QP}_\nk-\epsilon_\nk).
\end{equation}
After some simple derivations, the QP energy is calculated as~\cite{Shishkin2006}
\begin{equation}
\label{eq:GWQP}
\begin{split}
E^{\rm QP}_\nk =\epsilon_\nk + Z_\nk \mathrm{Re}
[ \bra{\psi_\nk} \hat{T}& + \hat{V}_{n-e} + \hat{V}_{H}\ket{\psi_\nk}\\
&+ \Sigma_\nk(\epsilon_\nk) -\epsilon_\nk],
\end{split}
\end{equation}
where $\hat{T}$  is the kinetic energy operator, $\hat{V}_{n-e}$ the nuclei potential,
$\hat{V}_{H}$  the Hartree potential, and $Z_\nk$ the renormalization factor given by
\begin{equation}
\label{eq:Z}
Z_\nk={\left(1-\left.\cfrac{\partial\mathrm{Re}[\Sigma_\nk(\omega)]}{\partial\omega}
\right|_{\omega=\epsilon_\nk}\right)}^{-1}.
\end{equation}
In principle, one could calculate the QP energies by searching the root of
equation
$  E^{\rm QP}_\nk=\mathrm{Re} [ \bra{\psi_\nk} \hat{T}+ \hat{V}_{n-e} + \hat{V}_{H}\ket{\psi_\nk}+ \Sigma_\nk(E^{\rm QP}_\nk)] $
numerically.
For solids, this does not make a sizeable difference in the QP energies compared to the linearization.
In the present work we therefore only show the calculated QP energies from the linearized version
to compare with the conventional implementation where the linearization was used as well.

The spectral functions are calculated as the imaginary part of the interacting Green's
function, which is  calculated from the Dyson-equation~\cite{MiyakePRB2013}
\begin{equation}
\label{eq:A}
\begin{split}
 A_{\nk}&(\omega) = \cfrac{1}{\pi}|\mathrm{Im}[G_{\nk}(\omega)]| \\
&=  \cfrac{1}{\pi}\cdot \cfrac{|\mathrm{Im}[\Delta \Sigma_\nk(\omega)]|}
{(\omega-\epsilon_\nk-\mathrm{Re}[\Delta\Sigma_\nk(\omega)])^2
+\mathrm{Im}[\Delta\Sigma_\nk(\omega)]^2}\,,
\end{split}
\end{equation}
where $\Delta \Sigma_\nk(\omega)=\bra{\psi_\nk} \hat{T} + \hat{V}_{n-e} + \hat{V}_{H}\ket{\psi_\nk}
+ \Sigma_\nk(\omega) -\epsilon_\nk $.

\begin{table*}
\caption{Positions of conduction band (CB) minimum at $\Gamma$ ($\Gamma_{c}$)
and $X$ ($X_{c}$), valence band (VB) maximum at $X$  ($X_{v}$)
with respect to the VB maximum at $\Gamma$, as well as the band gap.
Spin-orbit coupling (SOC) and finite basis-set corrections are not included.
The crystal structures, lattice constants,  and experimental band gaps are identical
to Ref.~\cite{Shishkin2007} and references therein. }
\begin{ruledtabular}
\begin{tabular}{lccccccccccc}
& \multicolumn{2}{c}{$\Gamma_{c}$}  &  \multicolumn{2}{c}{$X_{c}$} &  \multicolumn{2}{c}{$X_{v}$}  &  \multicolumn{3}{c}{band gap}  & Crystal  &  Lattice \\
 \cline{2-3}  \cline{4-5} \cline{6-7}  \cline{8-10}
& $G_0W_0r$ & $G_0W_0$ & $G_0W_0r$ & $G_0W_0$ &  $G_0W_0r$ &  $G_0W_0$ &  $G_0W_0r$ & $G_0W_0$ &  Expt.  & structure & constant ($\AA$)  \\
 \hline
Si   &   3.22 &   3.23 &   1.24 &   1.25 &  -2.89 &  -2.89   &   1.15 &    1.16 &    1.17 & diamond    & 5.430 \\
GaAs &   1.33 &   1.34 &   1.86 &   1.88 &  -2.79 &  -2.77   &   1.33 &    1.34 &    1.52 & zincblende & 5.648 \\
SiC  &   7.40 &   7.41 &   2.30 &   2.31 &  -3.36 &  -3.35   &   2.30 &    2.31 &    2.40 & zincblende & 4.350 \\
ZnO  &   2.10 &   2.06 &   6.73 &   6.66 &  -2.31 &  -2.28   &   2.10 &    2.06 &    3.44 & zincblende & 4.580 \\
C    &   7.39 &   7.39 &   6.07 &   6.08 &  -6.66 &  -6.66   &   5.49 &    5.50 &    5.48 & diamond    & 3.567 \\
BN   &  11.14 &  11.14 &   6.16 &   6.17 &  -5.28 &  -5.27   &   6.16 &    6.17 & 6.1-6.4 & zincblende & 3.615 \\
MgO  &   7.27 &   7.27 &  11.47 &  11.48 &  -1.55 &  -1.54   &   7.27 &    7.27 &    7.83 & rocksalt   & 4.213 \\
LiF  &  13.68 &  13.68 &  20.20 &  20.20 &  -1.21 &  -1.19   &  13.68 &   13.68 &   14.20 & rocksalt   & 4.010 \\
\end{tabular}
\end{ruledtabular}
\label{table:PBE_GW}
\end{table*}

\begin{figure*}
\begin{center}
\includegraphics[width=1.00\textwidth]{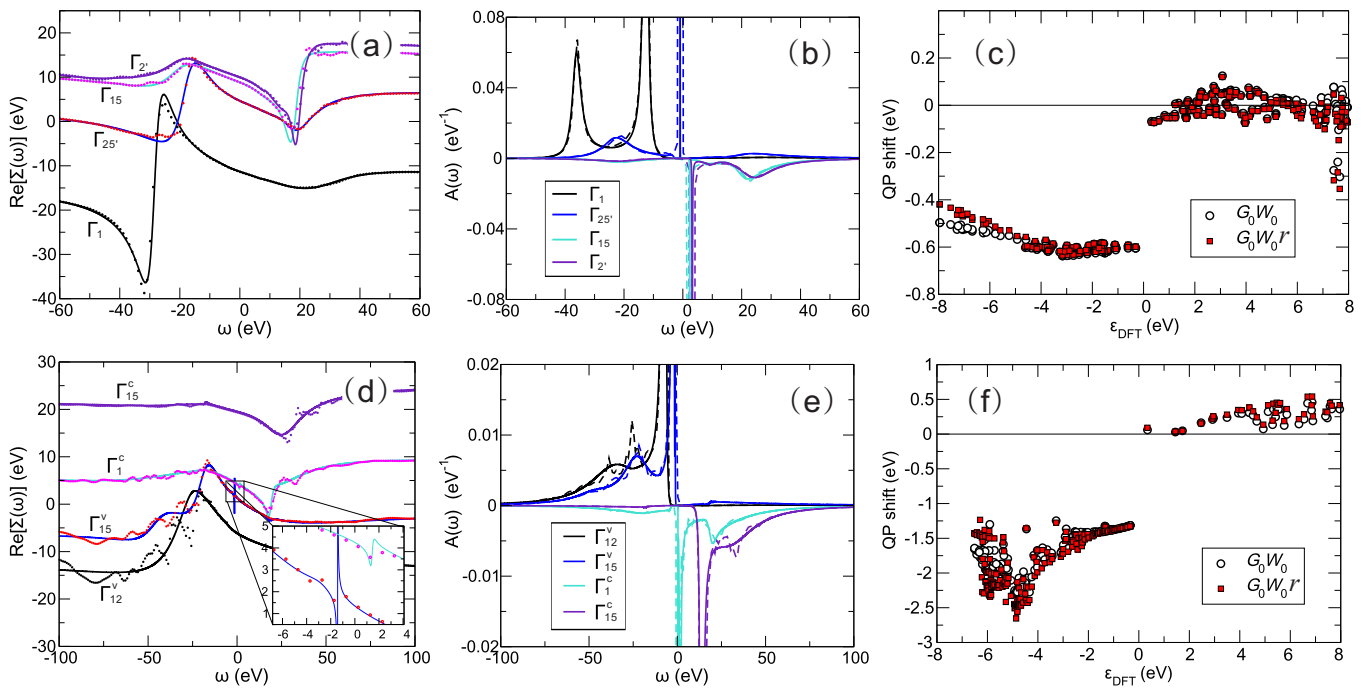}
\end{center}
\caption{(color online) The real part of the diagonal elements of the self-energy
Re[$\Sigma_{\nk}(w)$] [(a) and  (d)], and the spectral functions
$A_{\nk}(w)$ of the Green's functions at the $\Gamma$ point [(b) and  (e)], as well as the QP shift versus the
DFT eigenvalues [(c) and  (f)] for Si (first row) and ZnO (second row).
The solid lines and dotted (broken) lines in [(a),   (d),  (b) and  (e)], respectively,
 specify the results from $G_0W_0r$ and $G_0W_0$.
Note that the sign of the spectral functions for the unoccupied states in  (b) and  (e) is intentionally
reversed for clarity. The inset in  (d) shows the zoom-in plot for the local satellites.}
\label{fig:insulator_spectral}
\end{figure*}

\section{Technical details}\label{sec:tech}
Our low-scaling $GW$ scheme has been implemented in the Vienna
\emph{ab initio} simulation package (VASP)~\cite{KresseVASP1993, KresseVASP1996}.
For all the calculations presented here, the ultrasoft (US) PAW potentials with an appendix ({\tt $\_$GW})
released with VASP.5.2 were used unless otherwise explicitly specified.
These potentials are constructed by using additional
projectors above the vacuum level and thus describe well the high-energy scattering
properties of the atoms. The plane-wave cutoff for the orbitals was chosen to be the maximum
of all elements in the considered material. The energy cutoff for the response function
was chosen to be half of the plane-wave cutoff. To sample the Brillouin zone,
8$\times$8$\times$8 $k$-point grids centered at the $\Gamma$ point were used
except for Cu where the grids were increased to 10$\times$10$\times$10.
For the tested materials, the experimental lattice constants at low temperature
 (if available, otherwise at room temperature) were used. The total number
of bands was chosen to be 480, which is sufficient to obtain the converged QP energies
for most of the materials considered, except for GaAs and ZnO where the convergence is very slow.
It was suggested that thousands of orbitals are required for accurate predictions for ZnO~\cite{Shih2010},
but this finite-basis-set correction is beyond the scope of this work.
In fact, for the present setup the errors in some QP energies are large, with errors of, e.g., 0.5 eV for
ZnO. For more accurate results we refer to
the previous publication by some of the present authors~\cite{Jiri2014}.

Clearly, the purpose of the present work is not to basis set converge the calculations
(this is of course possible with the present implementation, as it was possible in the standard framework).
Instead, we restrict ourselves to validating the low-scaling $GW$
implementation by comparing the results with the already widely used conventional
$GW$ implementation. Hence, the same setups (crystal structure, potential, $k$ points and so on)
were used for both $G_0W_0r$ and $G_0W_0$ calculations.
In addition, finite basis-set corrections for QP energies discussed in Ref.~\cite{Jiri2014}
are not taken into account for neither $G_0W_0r$ nor $G_0W_0$ calculations.

The actual $GW$ calculations involve three steps: (i) A self-consistent KS-DFT
calculation was performed using the Perdew-Burke-Ernzerhof (PBE) functional~\cite{PBE1996}.
(ii) The one-electron wave functions and eigenenergies of all unoccupied (virtual)
orbitals spanned by the plane wave basis set were evaluated by an exact diagonalization
of the previously determined self-consistent KS Hamiltonian. (iii) The
$GW$ calculations were carried out. For all the  materials considered,
the number of imaginary time/frequency points in $G_0W_0r$ calculations
was set to 20, whereas the number of real frequency points was chosen to be 200
for $G_0W_0$ calculations. Increasing the number of grid points further changes
the QP energies by less than 0.01 eV.

\section{RESULTS }\label{sec:results}

\subsection{Results for semiconductors and insulators}\label{sec:insulator}
Table \ref{table:PBE_GW} shows the QP energies and band gaps for
the tested semiconductors and insulators predicted by $G_0W_0r$ and $G_0W_0$.
First, we emphasize that our $G_0W_0$@PBE results are consistent with previous
calculations~\cite{Shishkin2007}. As expected, the band gaps calculated by $G_0W_0$@PBE
are slightly underestimated compared to the experimental values.
Improvements further towards experimental gaps have been achieved either by
$GW_0$@PBE (iterating the one-electron energies only in $G$)~\cite{Shishkin2007},
or by $G_0W_0$@HSE (using the hybrid functionals as a starting point)~\cite{Fuchs2007}.
The best agreement with experimental values thus far has been achieved by $GW^{\text{TC-TC}}$
(self-consistent $GW$ with the vertex correction only in $W$)~\cite{ShishkinPRL2007}.
We note again that finite basis-set corrections~\cite{Jiri2014} have not been used here,
which would increase the gap for ZnO by 0.3-0.4 eV, for instance.
Second, one can see that the agreement between  the results from $G_0W_0r$
and  $G_0W_0$ is remarkably good,
validating our low-scaling $GW$ implementation. Specifically,
for the $sp$ semiconductors and insulators (Si, SiC, C, BN, MgO and LiF),
the difference in QP energies and band gaps between $G_0W_0r$ and $G_0W_0$ is
not larger than 0.02 eV. This is even true for GaAs with localized $d$-orbitals.
Except for ZnO the $G_0W_0r$ seems to have the tendency to yield a slightly smaller
downwards shift (below 0.02 eV) for valence and conduction bands compared
to $G_0W_0$.
However, for ZnO the difference of the calculated gaps between $G_0W_0r$ and $G_0W_0$
is larger (0.04 eV) since the self-energy exhibits many poles from $d$-$p$ excitations
at energies around $-$40 eV [see Fig. \ref{fig:insulator_spectral}(e) below].

To further assess our low-scaling $GW$ implementation, we plot the diagonal elements
of the self-energies and spectral functions at the $\Gamma$ point for some chosen bands
around the Fermi level, as well as the QP shift versus the
DFT eigenvalues for Si and ZnO in Fig. \ref{fig:insulator_spectral}.
The results obtained from the conventional $G_0W_0$ are also presented for comparison.
Overall, the agreement between the results from $G_0W_0r$ and $G_0W_0$ is very good,
in particular for the region close to the Fermi level. Specifically,
for Si, the self-energies and spectral functions (including the spectral
background and contributions from plasmons) calculated by $G_0W_0r$
agree nicely with the ones from $G_0W_0$ [see Figs. \ref{fig:insulator_spectral}(a) and \ref{fig:insulator_spectral}(b)].
This is achieved by employing Thiele's reciprocal difference method.
Solving for the Pad\'{e} coefficients directly, however, yields less satisfactory results (not shown here).
For ZnO, the agreement in the self-energies and spectral functions is still good.
Even the small satellites in bands $\Gamma_{15}^\text{v}$ and $\Gamma_1^\text{c}$ are reproduced [see the inset in
Fig. \ref{fig:insulator_spectral}(d)]. However, satellites far from the Fermi level
have been smoothed by the analytic continuation.

In contrast, there exist larger deviations in the region far away from the Fermi level.
As shown in Figs. \ref{fig:insulator_spectral}(c) and \ref{fig:insulator_spectral}(f),  the difference in
the QP shift between $G_0W_0r$ and $G_0W_0$ increases as the binding energies
increase above 4 eV. The reason can be easily understood.
Considering band $\Gamma_1$  of Si for instance, the QP peak is not sharp.
Instead, it is broadened with a width of around 5 eV,
as shown in Fig. \ref{fig:insulator_spectral}(b). Therefore, it is difficult to
obtain the exact position of the quasiparticle.
This is true for both  $G_0W_0r$ and $G_0W_0$.
In addition, the QP peaks measured from the angle-resolved photoelectron spectroscopy (ARPES)
would be as broad as in the $GW$ approximation so that the errors
are in fact negligible compared to the width of the peak.

\subsection{Results for metals}\label{sec:metal}

\begin{figure*}
\begin{center}
\includegraphics[width=1.00\textwidth]{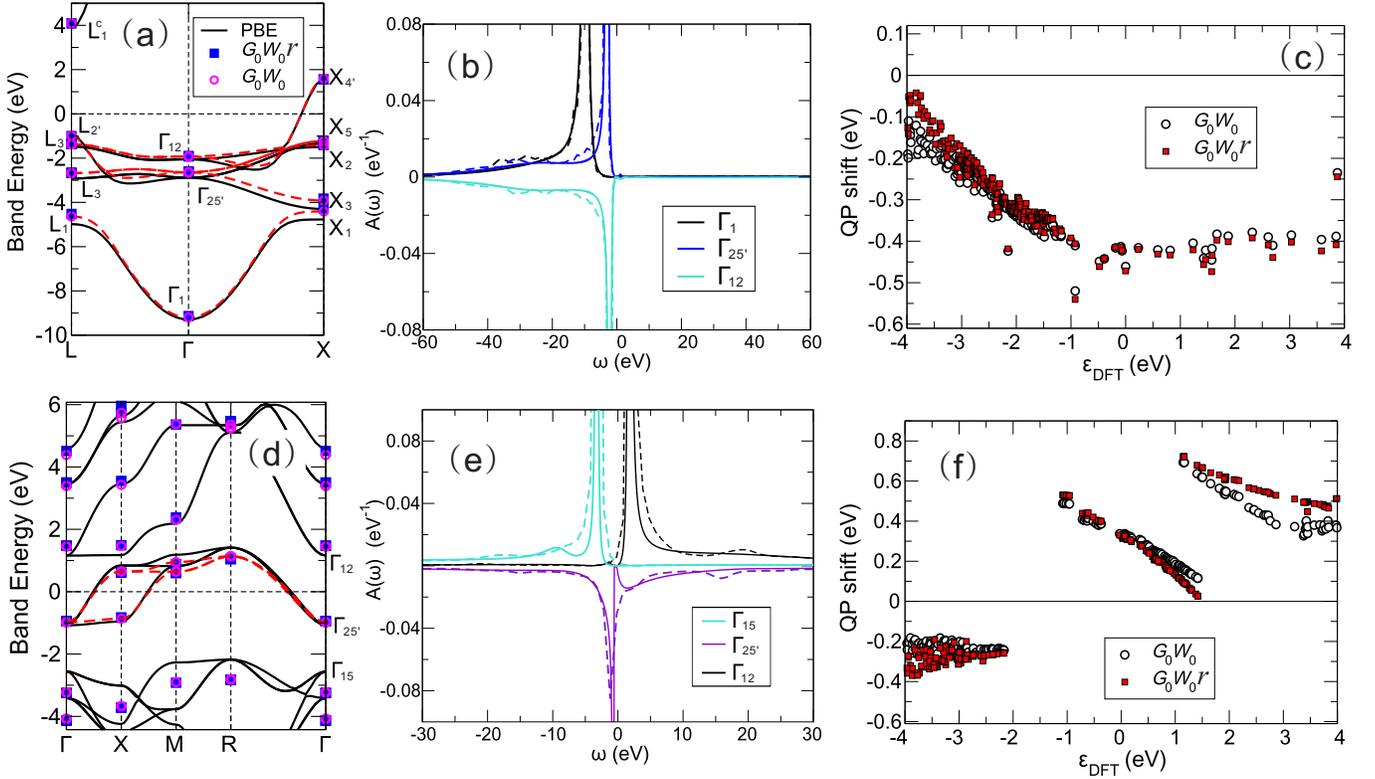}
\end{center}
\caption{(color online) Band structures [(a) and (d)], spectral functions $A_{\nk}(w)$ of the Green's functions
at the $\Gamma$ point [(b) and (e)], and QP shift versus the DFT eigenvalues [(c) and  (f)]
 for Cu (first row) and SrVO$_3$ (second row). Note that in  (a) and (d) the red broken lines
 specify the Wannier interpolated band structure from $G_0W_0$. PBE and $GW$ Fermi energies
 are aligned at zero.}
\label{fig:metal_spectral}
\end{figure*}

\begin{table*}
\caption{QP energies (eV) of Cu predicted by $G_0W_0r$ and $G_0W_0$ using the
norm-conserving (NC) GW PAW potential (Cu\_sv\_GW\_nc). Basis-set corrected $G_0W_0$ QP energies
are also given for comparison. The labeling of the high-symmetry points are shown in Fig. \ref{fig:metal_spectral}(a).
The results are compared to the pseudopotential plane wave (PPW) values obtained by Marini \emph{et al.}~\cite{MariniPRL2002}
and the full potential linear muffin-tin orbital (LMTO) calculations by
Zhukov  \emph{et al.}~\cite{Zhukov2003}. Experimental data are taken from Ref.~\cite{Courths1984}. }
\begin{ruledtabular}
\begin{tabular}{ccccccccc}
&   & PBE & $G_0W_0r$ & $G_0W_0$    & corrected $G_0W_0$ &  PPW~\cite{MariniPRL2002}  &  LMTO~\cite{Zhukov2003}   &  Expt.~\cite{Courths1984} \\
 \hline
\multirow{3}{2cm}{Positions of $d$ bands}
& $\Gamma_{12}              $ & $-2.05$ &       $-1.92$ & $-1.92$ & $-2.11$ & $-2.81$ & $-2.36$ & $-2.78$  \\
& $X_5                      $ & $-1.33$ &       $-1.22$ & $-1.23$ & $-1.45$ & $-2.04$ & $-1.63$ & $-2.01$  \\
& $L_3                      $ & $-1.47$ &       $-1.36$ & $-1.37$ & $-1.58$ & $-2.24$ & $-1.78$ & $-2.25$  \\
\multirow{4}{2cm}{Widths of $d$ bands}
& $\Gamma_{12}-\Gamma_{25'} $ & $ 0.84$ &       $ 0.70$ & $ 0.72$ & $ 0.69$ & $ 0.60$ & $ 0.81$ & $ 0.81$  \\
& $X_5-X_3                  $ & $ 2.97$ &       $ 2.61$ & $ 2.68$ & $ 2.60$ & $ 2.49$ & $ 2.92$ & $ 2.79$  \\
& $X_5-X_1                  $ & $ 3.44$ &       $ 3.05$ & $ 3.18$ & $ 3.10$ & $ 2.90$ & $ 3.37$ & $ 3.17$  \\
& $L_3-L_3                  $ & $ 1.44$ &       $ 1.30$ & $ 1.31$ & $ 1.26$ & $ 1.26$ & $ 1.43$ & $ 1.37$  \\
& $L_3-L_1                  $ & $ 3.51$ &       $ 3.16$ & $ 3.26$ & $ 3.16$ & $ 2.83$ & $ 3.42$ & $ 2.91$  \\
\multirow{2}{2cm}{Positions of $s/p$ bands}
& $\Gamma_1                 $ & $-9.29$ &       $-9.14$ & $-9.20$ & $-9.18$ & $-9.24$ & $-9.35$ & $-8.60$  \\
& $L_{2'}                   $ & $-0.92$ &       $-1.00$ & $-0.98$ & $-1.02$ & $-0.57$ & $-0.92$ & $-0.85$  \\
\multirow{1}{2cm}{$L$ gap}
& $L_1^c-L_{2'}             $ & $ 4.80$ &       $ 5.09$ & $ 5.08$ & $ 4.98$ & $ 4.76$ & $ 4.78$ & $ 4.95$  \\
\end{tabular}
\end{ruledtabular}
\label{table:QP_Cu}
\end{table*}

\begin{table}
\caption{Timings in minutes for $G_0W_0r$ and $G_0W_0$ calculations
for different bulk Si diamond supercells. The calculations were done for 64 QP energies
using the  $\Gamma$ point only and using the real valued
$\Gamma$ only VASP version. }
\begin{ruledtabular}
\begin{tabular}{ccccccc}
Atoms&    Cores    & \multicolumn{2}{c}{Time} & \multicolumn{3}{c}{Time$\times$cores/atoms$^3\times10^3$} \\
\cline{3-4} \cline{5-7}
     &   &$G_0W_0$  & $G_0W_0r$ & & $G_0W_0$  & $G_0W_0r$  \\
\hline
    16  &   16 &   9.18 &  2.50 && 35.86 &  9.78 \\
    24  &   20 &  18.94 &  4.14 && 27.40 &  5.99 \\
    36  &   48 &  41.68 &  5.65 && 42.88 &  5.82 \\
    54  &   64 & 104.53 & 12.07 && 42.49 &  4.91 \\
\end{tabular}
\end{ruledtabular}
\label{table:Timings_M}
\end{table}

\begin{figure}
\begin{center}
\includegraphics[width=0.38\textwidth]{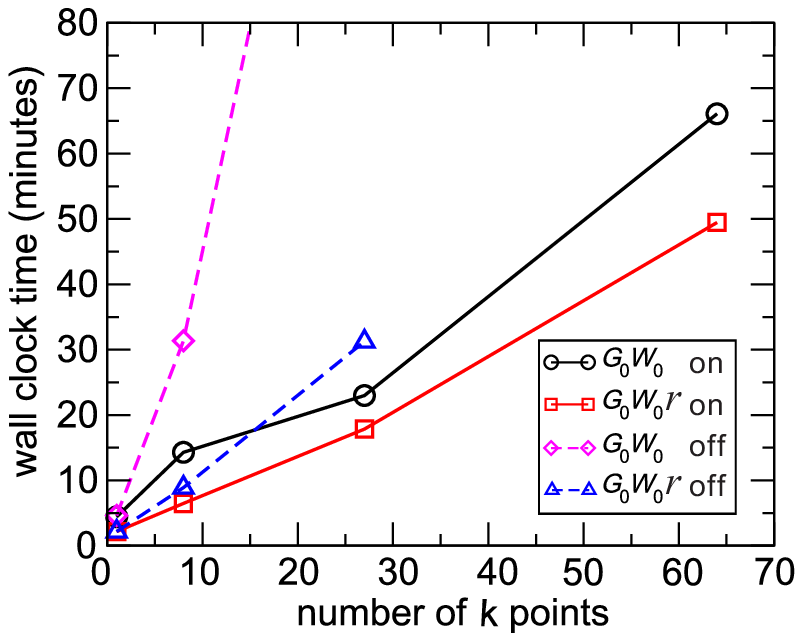}
\end{center}
\caption{(color online)
 Computational time for $G_0W_0r$ and $G_0W_0$ calculations with
symmetry switched on or off
on a bulk Si diamond supercell with 16 atoms as a function of the number
of $k$ points (in the full Brillouin zone). Note that the computational
time of $G_0W_0$ for the $3\times3\times3$ case without symmetry is about 162 minutes,
which is not shown in the figure.
The calculations are done for 64 QP energies using 64 cores.}
\label{fig:Timings_k}
\end{figure}

Now we turn to the QP calculations for metals where some extra considerations are required.
For metallic systems there exists a non-vanishing probability that an electron is excited
within one and the same band. These transitions are so-called intraband transitions and lead to the so
called Drude term for the long-wavelength limit ($\qq\rightarrow0$).
Following similar strategies as in Refs.~\cite{Gajdos2006,AmbroschDraxl2006} we derived the head of the intraband
dielectric function in the imaginary frequency domain,
\begin{equation}
\label{eq:Drude}
\varepsilon^{intra}_{\alpha\beta}(\ii\omega)=\cfrac{\bar\omega^2_{\alpha\beta}}{\omega^2}.
\end{equation}
Here the tensor $\bar\omega_{\alpha\beta}$  is the plasma frequency and its square
is defined as,
\begin{equation}
\label{eq:Plasmon}
\bar\omega^2_{\alpha\beta}=-\cfrac{4\pi e^2}{\Omega_C}\sum_\nk 2
\cfrac{\partial f(\epsilon_\nk)}{\partial \epsilon_\nk}
\left(\bf{e}_\alpha \cdot \cfrac{ \partial \epsilon_\nk}{\partial \kk}\right)
\left(\bf{e}_\beta  \cdot \cfrac{ \partial \epsilon_\nk}{\partial \kk}\right),
\end{equation}
where the factor of 2 is due to the spin-degenerate systems considered here, $\Omega_C$ is the volume of the
unit cell and $\bf{e}_\alpha$ is the unit vector along the Cartesian coordinate ${\alpha}$.
It should be noted that the intraband transitions are only non-vanishing
for the head of the dielectric functions. For the wings and body they are both zero.

As a test, we calculated the QP energies for the metals Cu and SrVO$_3$
and compare the results with the ones from the conventional $G_0W_0$
in Fig. \ref{fig:metal_spectral}(a) and \ref{fig:metal_spectral}(d). To guide the eye,
the PBE band structures and interpolated $G_0W_0$ QP band structures
obtained with the wannier90 code~\cite{Marzari1997, Souza2001} are also displayed.
One can see that good agreement between $G_0W_0r$ and $G_0W_0$
is achieved for both Cu and SrVO$_3$, indicating that our low-scaling
$GW$ implementation is robust and applies to metals as well.
However, if we take a closer look at the spectral functions,
as shown in Figs. \ref{fig:metal_spectral}(b) and \ref{fig:metal_spectral}(e), we observed that although
the main QP peaks are well reproduced, the plasma and some satellites
are again smoothed by the analytic continuation.

Figures \ref{fig:metal_spectral}(c) and  \ref{fig:metal_spectral}(f) further show the QP shift versus
DFT eigenvalues for Cu and SrVO$_3$, respectively.  For the noble metal
Cu, in the energy region of the plot, overall, the QP shift difference between  $G_0W_0r$
and $G_0W_0$ is not exceeding 0.1 eV.  Analogous behavior as for Si is observed for
Cu. The further one moves away from the Fermi level, the larger is the QP shift difference. This is due to
the large broadening of the QP peak for the corresponding bands.
This is also true for the metal SrVO$_3$.
The negative slope of the QP shift between $-1.5$ and 1.5 eV implies a shrinking of the $t_\text{2g}$
bands as compared to the DFT results, which was observed in other $GW$ studies as well~\cite{MiyakePRB2013,Tomczak2016}.
In the region far away from the Fermi level QP differences are visible, but
the maximum difference is smaller than 0.2 eV.

In Table \ref{table:QP_Cu}, we show in detail the QP energies of Cu predicted by $G_0W_0r$
and $G_0W_0$ and compare our results with other theoretical calculations and experiment.
The most significant error in the PBE one-electron energies is the wrong description
of the absolute positions and the bandwidth of the $d$ bands. For instance,
the highest $d$ band at $X_5$ is located at $-1.33$ eV in PBE, 0.68 eV above
the experimental value of $-2.01$ eV~\cite{Courths1984}.
The bandwidth of the $d$ bands is widened compared to the experiment (see $X_5-X_1$).
Unfortunately though, our $G_0W_0$ does not improve the results significantly.
It shrinks the bandwidth of the $d$ bands towards the experimental values, but
predicts worse positions for the $d$ bands than PBE. However,
as already mentioned basis set errors might be substantial for the 480
bands employed in the present case.  To improve the results, finite basis-set corrections
were used as discussed in Ref.~\cite{Jiri2014}.
Indeed, with these corrections the absolute positions of the $d$ bands are lowered by about 0.2 eV.
Agreement with the full potential LMTO method~\cite{Zhukov2003}
is then reasonable. However, our QP $d$-band energies are still
way above those of  Marini \emph{et al.}~\cite{MariniPRL2002}.
We are pretty confident that the good agreement of these calculations with experiment
is largely fortuitous: the applied pseudopotentials somehow
canceled the errors introduced by the $G_0W_0$ approximation.
Compared to the LMTO data, we note that all our QP energies
are shifted upwards by 0.2 eV (except for $L_{2'}$). Of course,
our QP energies are reported with respect to the $G_0W_0$
Fermi energy, whereas, Ref.~\cite{Zhukov2003}
does not mention how and whether the Fermi energy was determined at the $G_0W_0$ level.
Using the DFT Fermi energy would  improve agreement
with Ref.~\cite{Zhukov2003}. For the widths of the
$d$ bands the present results are in very good agreement
with experiment, though, slightly improving upon the LMTO
data, which were generally above the experimental data.

We feel that the residual errors compared to experiment are to be expected and arise from
(i) the neglect of self-consistency (the DFT $d$ orbitals of Cu
are most likely too strongly hybridized with the $sp$ states),
and (ii) spurious self-interactions in the $GW$ approximation.
The latter error  can  be only eliminated via the
inclusion of vertex corrections in the self-energy.
Indeed, the importance of vertex corrections has been highlighted
for predicting the ionization potentials and $d$-electron binding
energies of solids~\cite{Gruneis2014}, with typical corrections for
the $d$ bands of 0.7~eV.

\subsection{Time complexity for large systems}

In order to investigate the scaling with respect to the system size
in our new implementation, we performed $G_0W_0r$  calculations
on different bulk Si diamond supercells with 16, 24, 36,  and 54 atoms using the
$\Gamma$ point only. For comparison, similar calculations have been
done for the conventional $G_0W_0$ code. Our $G_0W_0r$ implementation
displays clearly a better than  cubic scaling in the system size, as shown in Table \ref{table:Timings_M}.
The reason for this good scaling is that the contraction steps such as $GG$ and $GW$ scale
only quadratically in system size, and for the number of atoms considered here, construction
of the Green's function and manipulations of the self-energy matrix, which scale
profoundly cubically, are not yet dominating the total compute time. Furthermore, it needs
to be mentioned that the $G_0W_0r$ compute time includes the
calculation of the full non-diagonal self-energy at all frequency points (including {\em all}
off-diagonal elements), whereas the $G_0W_0$ code calculates
only few diagonal elements of the self-energy for the occupied and some unoccupied states.
Despite this the $G_0W_0r$ code substantially outperforms the
older $G_0W_0$ code.  Concerning scaling, the old  $G_0W_0$  code shows a slightly
less beneficial scaling, nevertheless  it is also closer to cubic than quartic in system size.
This relates to the fact that the quartic part (construction of polarizability
and self-energy in orbital basis) is done using high efficiency BLAS
level 3 calls, and hence this part becomes only dominant for very large systems,
typically beyond 100 atoms.

To test the scaling with respect to the number of $k$ points, we performed calculations
on a bulk Si diamond supercell with 16 atoms using 64 cores.
We note that the new code does not yet perform optimally if the number
of cores exceeds the number of atoms. This and the need to use a complex code version
explain why the timings for a single $k$ point in Fig.  \ref{fig:Timings_k} are hardly better
than for 16 cores shown in Table. \ref{table:Timings_M}.
As shown in Fig. \ref{fig:Timings_k}, the computational demand
increases almost perfectly linear in the number of $k$ points for $G_0W_0r$.
The slight deviation for the $4\times4\times4$ case arises from the need to pick a less
efficient parallelization strategy for this $k$-point set to be able to perform the calculation using the
memory available on 64 cores.
In contrast, $G_0W_0$ shows a roughly quadratic scaling in the number of $k$ points.
The bad scaling of the old code is, however, somewhat masked by its
efficient handling of symmetry. The old implementation uses small point group operations compatible with the considered
momentum transfer $q$, whereas the new code uses yet no symmetry when contracting
$GG$ or $GW$. Concomitantly, if symmetry is switched off, the new code becomes
only slower by a factor 2 for the $3\times3\times3$ $k$ points, whereas the
time for the old code increases to 162 minutes (off the scale, see the blue and pink broken lines in Fig. \ref{fig:Timings_k}).
Therefore, one would expect that the  $G_0W_0r$  code outperforms
 the old $G_0W_0$ code, in particular, if large low-symmetry unit cells are used and/or
if many $k$ points are used. It is however also clear that the old
code can be competitive or superior for small high-symmetry unit cells,
even if many $k$ points are used to sample the Brillouin zone.
For instance, for a cubic diamond unit cell or for fcc Cu,
the old code is usually much faster than the new $GW$ code.

\section{Conclusions}
In conclusion, we present a promising low-scaling $GW$ implementation within the PAW method,
which allows for fast QP calculations with a scaling that is roughly cubic in the system
size and linear in the number of $k$ points used to sample the Brillouin zone.
All implementation details have been given.
We apply the method to predict the quasiparticle energies and spectral functions
for typical semiconductors, insulators, and metals.
Comparison of the results with the ones from conventional $GW$ calculations shows
a good agreement between the two implementations.
Specifically, for semiconductors and insulators the positions of the bands and
the band gaps agree within $0.02$~eV for all considered materials except for ZnO.
Due to the low scaling of our new $GW$ implementation, we believe that
our method has great potential for applications, in particular for large unit cells.
In addition, our $GW$ self-energies are obtained in imaginary time/frequency
domain, which will facilitate an elegant combination of $GW$ with DMFT,
enhancing the predictive abilities of $GW$+DMFT for large correlated systems.
Finally, we have shown that with typical compute times around 12 minutes
on 64 cores for 54 silicon atoms,
$GW$ calculations are becoming a commodity.
We believe this will greatly
help to establish methods beyond density functional theory
in the realm of materials modeling.

\begin{acknowledgments}
This work was supported by the China Scholarship
Council (CSC)-Austrian Science Fund (FWF) Scholarship Program and
FWF within the SFB ViCoM (Grant No. F 41) and I597-N16
(research unit FOR 1346 of the Deutsche Forschungsgemeinschaft
and FWF).
Supercomputing time on the Vienna
Scientific cluster (VSC) is gratefully acknowledged.
JK is supported by the European Union's Horizon 2020 research and innovation
programme under the Marie Sklodowska-Curie grant agreement No 658705.
\end{acknowledgments}

\bibliographystyle{apsrev}
\bibliography{reference} 

\end{document}